# Trapping of a microsphere pendulum resonator in an optical potential

J. M. Ward,[1,2] Y. Wu (邬宇强),[2,3] V. G. Minogin,[1,3,4] and S. Nic Chormaic[2,3]

[1]*Dept. Applied Physics and Instrumentation, Cork Institute of Technology, Cork, Ireland*

[2]*Photonics Centre, Tyndall National Institute, Prospect Row, Cork, Ireland*

[3]*Physics Department, University College Cork, Cork, Ireland*

[4]*Institute of Spectroscopy Russ. Ac. of Sciences, 142190 Troitsk, Moscow Region, Russia*

We propose a method to spatially confine or corral the movements of a micropendulum via the optical forces produced by two simultaneously excited optical modes of a photonic molecule comprising two microspherical cavities. We discuss how the cavity enhanced optical force generated in the photonic molecule can create an optomechanical potential of about 10 eV deep and 30 pm wide, which can be used to trap the pendulum at any given equilibrium position by a simple choice of laser frequencies. This result presents opportunities for very precise all-optical self-alignment of microsystems.

The study of optical forces on micromechanical systems and coupled microcavities is of ever increasing importance as the size of photonic devices gets smaller and, therefore, optical forces acting on such devices become ever more significant [1-8]. The optical forces result from the dipole interaction created by electromagnetic fields in closely-spaced dielectric structures; light travelling in the structure leads to a polarization of the electric field of atoms in the material, thereby creating a dipole between the atoms and the light. When two such structures are brought close together the electric dipoles interact, creating a force the magnitude of which is proportional to the intensity of the propagating light. Momentum conservation arguments have been used to show that the average electromagnetic force acting on a body is equal to the surface integral of the time averaged Maxwell tensors [8].

Such optical forces can be viewed as efficient tools for controlling the motion of micro and nanobodies in space, including trapping and corralling small dielectric bodies such as planar slabs or waveguides [1], microdisk or microspherical resonators [2], and dielectric cavities of more complicated structure [3-5]. For example, in a recent study [8] on the optical forces on two planar waveguides excited by symmetric and anti-symmetric electromagnetic modes, the maximum achievable deflection for each of the waveguide's modes was discussed. Another proposal [6] was based on the spectral bonding and self-stabilization of two slabs of dielectric material, guiding a superposition of two transverse evanescent fields, thereby creating an "optical spring".

One important configuration for creating sufficiently large optical forces could consist of two closely positioned evanescently-coupled microresonators. In such an optomechanical system, the optical forces arise due to the excitation of symmetric and anti-symmetric electromagnetic modes with frequencies depending on the distance between the cavities [1-8]. In [1] the optical force between two microspheres was evaluated as a function of the frequency splitting, which is, in turn, a function of the separation of the two spheres. It was shown that a force of 100 nN could be achieved for two 32.4 $\mu$m spheres, separated by 500 nm, with Q factors of $10^8$, and pumped at 1mW. The authors considered the spring constant of the fiber stem onto which the sphere was attached and showed that, for a spring constant of 0.004 N/m, the displacement of the spheres due to optical forces could be as large as 1.5 $\mu$m. They also showed that identical size matching of the spheres during fabrication was not essential by considering the effects of differing values of *l* (polar mode number) and *m* (azimuthal mode number) for the two spheres. Van der Waals forces and electrostatic forces were also considered and these were found to be negligible and distinguishable, respectively, from the optical force.

Frequency splitting of high *Q* modes in systems consisting of two microresonators has already been extensively discussed [9-11] and experimentally observed for different shaped dielectric bodies [12-15]. Such a splitting arises from the degeneracy breaking of the co-resonant modes in the coupled microresonators. This results in symmetric and anti-symmetric modes being established in the microresonators in what can be viewed as the optical analogue to the bonding and anti-bonding modes in conventional molecules. When the symmetric mode of the coupled microresonators is excited an attractive force is generated that tends to pull the microresonators together. In contrast, the excitation of the anti-symmetric resonance leads to the generation of a repulsive force that pushes the microresonators apart. In general, each individual microresonator possesses a large set of optical resonances and each of these resonances is split in a coupled optomechanical system (for a perfect sphere).

In recent work [7], it was shown that the trapping optical potentials in a two-body optomechanical system can be created by exploiting the forces produced by the symmetric and anti-symmetric modes of *two neighboring optical resonances*. The system is assumed to be pumped by laser light slightly blue detuned from the crossover frequency of a symmetric and anti-symmetric mode. Such an approach allows for the creation of an attractive and repulsive force, the sum of which may form a potential well a few tens of eV deep [7]. This work focused on small micro-rings with radii of 2.5 $\mu$m, with a large free spectral range (4.5 THz) and a tuning range of 45 THz, so that crossings between the symmetric resonance of one whispering gallery mode (WGM) and the anti-symmetric resonance of a neighboring WGM can be achieved. If so, one laser line may be used to simultaneously excite the symmetric and anti-symmetric resonance at one of these crossings.

Here. we propose an alternative approach to spatially confine the motion of coupled microresonators. In our approach, the optical force is assumed to be produced by the symmetric and anti-symmetric modes of *the same optical resonance*. This is achieved by simultaneous excitation of the system by two-frequency laser light so that one monochromatic component of the light excites the symmetric mode and the second component excites the anti-symmetric mode. Specifically, we consider an optical system consisting of two microspheres [16, 17], one of which is evanescently coupled to an optical fiber carrying the exciting laser field and the other is free to move in space and can be viewed as a micropendulum. We show that, in our scheme, the movable microsphere can be localized in a region as small as 10 pm. An important feature of our approach is the opportunity it offers to trap the pendulum at *any* given spatial position,

since - by a judicious choice of the frequencies of the laser light - one can predefine the equilibrium position of the pendulum. We also stress that, here, the symmetric and anti-symmetric modes can be simultaneously excited by two fixed laser lines so that the pendulum may be spatially trapped at *any* frequency position where the splitting occurs, without size restrictions. One important application may be in the precision control of the gap between resonators of a photonic molecule, thus allowing for extremely fine tuning of the optical modes.

A schematic of the proposed approach is shown in Fig. 1. A photonic molecule, consisting of two microspheres, is optically pumped by two-frequency laser light using a tapered optical fiber. The first microsphere (Sphere 1) is fixed and unable to move, while the second, movable microsphere (Sphere 2) is suspended by a thin fiber stem and acts as a pendulum in the photonic molecule system. The length of the stem can be chosen in order to ensure that the pendulum has the desired spring constant. The microspheres are positioned so that they are separated by less than 500 nm to ensure that whispering gallery modes excited in the first microsphere are coupled into the second microsphere.

A qualitative description of the mechanical properties of the proposed system is shown in Fig. 2. Figure 2(a) depicts the symmetric and anti-symmetric modes of the photonic molecule originating from a degenerate mode with frequency, $v_0$. We assume that two-frequency laser light excites the symmetric and anti-symmetric modes of the photonic molecule. Laser light with frequency $v_1 > v_0$ effectively excites the anti-symmetric mode at a sphere separation, $x_1$. Light with frequency $v_2 < v_0$ excites the symmetric mode at a sphere separation, $x_2$. Generally, these two modes create a force on Sphere 2 at any position, with zero force at the equilibrium position, $x_0$, located between points $x_1$ and $x_2$, as shown in Fig. 2 (b). Specifically, when the movable microsphere, Sphere 2, is located at a relatively small distance around $x_1$ in the region $x < x_0$, it is mostly excited by the anti-symmetric mode with frequency close to $v_1$ and, accordingly, it is acted on by a positive force which attracts the microsphere to the equilibrium position from the left. In a similar way, if Sphere 2 is positioned in the region $x > x_0$ around $x_2$ it is mostly excited by the symmetric mode with frequency close to $v_2$ and is acted on by a negative force which attracts the microsphere to the equilibrium position from the right. Applying two-frequency laser light, therefore, generates an optical force that is responsible for the creation of a potential well with a minimum at $x_0$ as shown in Fig. 2(c). The position of the equilibrium point may be controlled by varying the laser frequencies, $v_1$ and $v_2$. Any disturbance of Sphere 2 from its equilibrium position could be detected as a shift in the resonance positions of the symmetric and anti-symmetric modes, and could be monitored using a very weak probe laser.

Assuming that the movable Sphere 2 is brought into the evanescent field of the fixed Sphere 1, we first evaluate the splitting of the optical resonance defined by a polar mode number $l$, azimuthal mode $m = l$ and radial mode $n = 1$. For a given resonance, the frequencies of the anti-symmetric, $v_a$, and symmetric, $v_s$, modes as a function of the separation between the microspheres can be found from [18,19]

$$v_{a,s} = \left(1 + \frac{(\beta \mp \alpha)}{2}\right) v_0, \qquad (1)$$

where $v_0$ is the unshifted modal frequency, $\alpha = \varsigma_\alpha l^{-3/2} e^{-x/d}$ is the frequency shift due to overlapping of the evanescent fields of the microspheres, and $\beta = \varsigma_\beta l^{-3/2} e^{-2x/d}$ is the frequency shift caused by the change in the optical path length of one microsphere due to the presence of the other. In the above equations, $d = \lambda/2\pi\sqrt{n_s^2 - 1}$ is the evanescent field decay length defined by the refractive index, $n_s$, of the microspheres at wavelength $\lambda = c/\nu$. $\varsigma_\alpha$ and $\varsigma_\beta$ are functions of microsphere dimensions and refractive indices. Specifically, we choose microspheres of radius, $R = 12.5\ \mu m$, made of fused silica with a refractive index, $n_S = 1.46$, and a quality factor, $Q = 10^8$. Each sphere is assumed to support an electromagnetic mode with numbers $l = 65$, $m = l$, $n = 1$, and a corresponding wavelength $\lambda = 1568$ nm or size parameter 50.0668, calculated according to the asymptotic expansion from [20]. For such microspheres, the free spectral range $\Delta \nu = c/2\pi R n_s = 2.6$ THz, $d = 234.6$ nm, $\varsigma_\alpha = 1.4$, and $\varsigma_\beta = 0.1$ [19].

In the proposed system each microsphere experiences an optical force defined by the optical energy stored in the microsphere. The optical force on Sphere 2 can be evaluated from

$$F = F_a + F_s, \qquad (2)$$

where the partial forces due to the anti-symmetric and symmetric modes are defined by [1,5]

$$F_{a,s} = -\frac{1}{\omega_{a,s}} U_{a,s} \frac{d\omega_{a,s}}{dx} = -\frac{1}{\omega_0 + \Delta\omega_{a,s}} U_{a,s} \frac{d(\omega_0 + \Delta\omega_{a,s})}{dx} \approx -\frac{1}{\omega_0} U_{a,s} \frac{d\Delta\omega_{a,s}}{dx} \qquad (3)$$

$U_a$ and $U_s$ are the energies of the electromagnetic field stored in the anti-symmetric and symmetric modes pumped by external fields with frequencies $\omega_1 = 2\pi\nu_1$ and $\omega_2 = 2\pi\nu_2$, such that

$$U_{a,s} = U_0 \frac{(\gamma/2)^2}{(\omega_{1,2} - \omega_{a,s})^2 + (\gamma/2)^2} \tag{4}$$

In Eq. (4) $\gamma$ is the linewidth of the cavity resonance, $U_0$ is the maximum energy stored in a single microsphere at resonance, such that $U_0 = PQ/\omega_0$. $Q = \omega_0/\gamma$ is the quality factor of a single microsphere, $P$ is the optical power, and $\omega_0 = 2\pi\nu_0$. Here, the symmetric and anti-symmetric modes are treated as individual Lorentzian shaped resonances and the energy in the photonic molecule varies with the detuning from each mode [15, 19, 21, 22].

The frequencies of the split modes for the microspheres are plotted in Fig. 3(a), for an intersphere gap of $250 \pm 0.07$ nm. The numerical results presented in Figs. 3(b) and 3(c) were obtained by choosing the equilibrium position of Sphere 2 to be $x_0 = 250$ nm. At equilibrium position, frequencies for the anti-symmetric and symmetric modes, $\nu_{0a} = \nu_a(x_0)$ and $\nu_{0s} = \nu_{0s}(x_0)$, are determined from Eq. (1). To obtain a zero value for the optical force at $x_0$, $\nu_1$ and $\nu_2$ are detuned from $\nu_{0a}$ and $\nu_{0s}$ by equal amounts. Detunings of $\delta = 0.7\gamma$, $3\gamma$ and $5\gamma$ are used for the results presented. Figure 3(b) shows the optical force on the movable microsphere, calculated for three different sets of applied laser frequencies, $\nu_1$ and $\nu_2$, and an optical power, $P = 1$ mW. Figure 3(c) plots the optical potential for the movable microsphere generated by the optical force determined from Eq. (2). The asymmetry in the potential well is due to the asymmetry between the attractive and repulsive forces. One can see that, as the detuning increases from $0.7\gamma$ to $5\gamma$, the depth of the potential well increases from 5 to 13 eV and the trap width increases from 10 to 58 pm.

Next, we evaluate the mechanical properties of a pendulum composed of the movable microsphere attached to a cantilever. In the geometry we consider, the oscillation frequency of the pendulum along the $x$-direction is defined by the spring constant, $\kappa_{opt}$, of the optical force and the spring constant, $\kappa_c$, of the cantilever. Assuming that $\kappa_c < \kappa_{opt}$ and the mass of the microsphere, $M$, exceeds that of the cantilever, $m$, we can evaluate the oscillation frequency as $\Omega_{opt} = \sqrt{\kappa_{opt}/M}$. As an example, we evaluate the spring constant of the optical force for a silica microsphere with radius, $R = 12.5\,\mu\text{m}$, and mass, $M = 1.8\times 10^{-11}$ kg (at a density of $\rho = 2200\,\text{kg/m}^3$) and assuming a detuning, $\delta = 3\gamma$ (c.f. Fig. 3(b)). For this case, the width of the trapping region is $\Delta d \approx 30\,\text{pm}$. The spring constant is evaluated from Eqs. (1)-(4) and is given by $\kappa_{opt} = -(dF/dx)_{x=x_0} = 4500$ N/m and the oscillation frequency is given by $\Omega_{opt}/2\pi = 2.5\,\text{MHz}$.

It is worth noting that the quantities $\kappa_{opt}$ and $\Omega_{opt}$, produced by the optical force, considerably exceed the corresponding values defined by the material properties of the cantilever, $\kappa_c$ and $\Omega_c$. The spring constant of a cylindrical cantilever of radius, $a$, and length, $L$, defines the deflection oscillation mode of the pendulum. This can be evaluated using $\kappa_c = (3\pi/4)(Ea^4/L^3)$, where $E$ is Young's modulus. For a system fabricated in our laboratory, composed of a silica cantilever of radius, $a = 1\,\mu\text{m}$, and length, $L = 0.01\,\text{m}$, attached to a silica microsphere of radius, $R = 12.5\,\mu\text{m}$ and mass, $M = 1.8\times 10^{-11}$ kg, we obtain $\kappa_c = 1.7 \times 10^{-7}$ N/m and $\Omega_c/2\pi = \left(\sqrt{\kappa_c/M}\right)/2\pi \approx 15$ Hz. The extremely small value of $\kappa_c$ ensures that, in the absence of the optical force, the microsphere at the end of the pendulum is essentially free to move in the horizontal plane. For silica, we have taken Young's modulus, $E = 72.6$ GPa, and the density, $\rho = 2200$ kg/m$^3$.

Taking into account the spring constant of the pendulum along the *x*-direction due to the optical force one can evaluate that, at a temperature, $T$, the uncertainty in the pendulum position is $\Delta x = \sqrt{2k_B T/\kappa_{opt}}$. For a detuning, $\delta = 3\gamma$, when $\kappa_{opt} = 4500$ N/m, the position uncertainty at room temperature, $T = 300$ K, is estimated to be as low as 1.4 pm. The position uncertainty can be further reduced by using a larger spring constant and lower temperature. However, a larger spring constant would also reduce the maximum displacement of the pendulum that can be achieved by the optical force and the sensitivity of the pendulum to external forces.

In addition to the dramatic influence on the deflection mode along the *x*-direction, the optical force also influences the deflection mode along the orthogonal direction, i.e. the *y*-axis (in and out of the plane in Fig.1). If a disturbance causes a small displacement, $\delta y$, of the pendulum along the *y*-direction, the optical force produces a restoring force component, $F_y = F(\delta y/2R)$, along the *y*-direction. Due to thermal fluctuations the optical force has a typical value of about $F = \kappa_{opt}\Delta x$ and, accordingly, the restoring force along the *y*-axis can be characterized by a spring constant of about $\kappa_y = \kappa_{opt}(\Delta x/2R) = 2.5\times 10^{-4}$ N/m. This optical spring constant exceeds the material spring constant, $\kappa_c$, by three orders of magnitude. Accordingly, the oscillation frequency along the *y*-direction is defined also by the optical force, and has a value of ~ $\Omega_y/2\pi = \left(\sqrt{\kappa_y/M}\right)/2\pi \approx 0.6$ kHz. The uncertainty in the pendulum position along the *y*-axis can be evaluated by $\Delta y = \sqrt{2k_B T/\kappa_y} = 5.8$ nm.

Until now, we have discussed parameters of the optomechanical pendulum for the deflection modes that cause oscillations of the pendulum in the horizontal plane. In general, one should consider the influence of the optical forces on compression and torsion modes of the pendulum. In what follows, we show that the contribution to both of these modes from optical forces is negligibly small.

For the case of a compression mode, the mechanical oscillations of the considered pendulum in the vertical direction, i.e. the $z$-direction, are defined by the spring constant, $\kappa_z = \pi a^2 E / L$, where, as before, $a$ is the radius and $L$ is the length of a silica cantilever, and $E$ is Young's modulus for silica. The associated vertical oscillation frequency of the microsphere pendulum along the $z$-direction is defined by its material properties and is given by $\Omega_z = \sqrt{\kappa_z / M}$. Using the aforementioned values for $a$, $L$ and $M$ we obtain $\kappa_z = 23 \, \text{N/m}$ and $\Omega_z / 2\pi = 180 \, \text{kHz}$. The spring constant due to the vertical component of the optical force can be evaluated in a manner similar to the evaluation of the spring constant due to the $y$-component of the optical force. If, due to any reason, the movable microsphere is shifted in the vertical direction by an amount $\delta z$ this produces a vertical component of the optical force, $F_v$, such that $F_v \approx F_{opt}(\delta z / 2R) = \kappa_v \delta z$, where $\kappa_v = F_{opt}/2R \approx \kappa_{opt}(\Delta x/2R) = \kappa_y = 2.5 \times 10^{-4} \, \text{N/m}$. Contrary to the situation in the $y$-direction, the friction coefficient, $\kappa_v$, due to the optical force in the vertical direction is very small compared to $\kappa_z = 23 \, \text{N/m}$, due to the mechanical properties of the pendulum and can, therefore, be neglected. For the vertical oscillation amplitude we obtain, at room temperature, $\Delta z = \sqrt{2 k_B T / \kappa_z} = 19 \, \text{pm}$. The oscillation amplitude is very small when compared to the radius of the microsphere, i.e. $\Delta z / R \approx 10^{-6}$; therefore, we deduce that the vertical displacement of the movable microsphere does not influence the optical oscillations of the pendulum along the *x*-direction.

For a torsion mode, the spring constant, $\kappa_\varphi$, due to the material properties of the pendulum is defined by the shear modulus of the silica cantilever, $G = 31\,\text{GPa}$, the cantilever radius, $a = 1\,\mu\text{m}$, and the length of the cantilever, $L = 0.01\,\text{m}$, such that $\kappa_\varphi = \pi a^4 G / 2L = 4.9 \times 10^{-12}\,\text{Nm}$. The corresponding torsion oscillation frequency is defined by $\kappa_\varphi$ and the moment of inertia of the microsphere, $I = \tfrac{2}{5} MR^2 = 1.1 \times 10^{-21}\,\text{kg}\,\text{m}^2$, as $\Omega_\varphi/2\pi = \left(\sqrt{\kappa_\varphi/I}\right)/2\pi \approx 10.6\,\text{kHz}$. The optical force cannot influence the torsional motion due to the angular symmetry of the evanescent field. Fluctuations of the angular position of the pendulum due to thermal torsional motion can be evaluated as $\Delta\varphi = \sqrt{2k_B T / \kappa_\varphi} = 4.1 \times 10^{-5}$ rad. Again, the extremely small value of the torsion fluctuation angle shows that the torsional motion of the pendulum cannot influence optical oscillations along the *x*-direction.

In conclusion, we have proposed a method to trap and corral an optomechanical system consisting of two microspheres, one fixed and one free to move in space, i.e. a microsphere pendulum. The proposed approach relies on a two-frequency laser field, thereby allowing one to restrict the motion of the cavity pendulum inside a very small region of space at any given equilibrium position. We have shown that, for a $3\gamma$ detuning of the two-frequency laser field, the optical potential could be 10 eV deep and 30 pm wide. The room temperature energy of the pendulum is ~25 meV and, in the absence of the two-frequency laser field, the thermal displacement of the pendulum, $\Delta x = 0.147\,\mu\text{m}$. However, when the two-frequency laser field is applied, the optical restoring force, $\kappa_{opt}$, of the trap could be as large as 4500 N/m, thus restricting the thermal displacement to 1.4 pm. To move the pendulum out of the trap it would need to be displaced by $\pm 15$ pm with a force greater than 150 nN; such as disturbance would be observable as a change in the splitting frequencies. Control of the equilibrium position of the microsphere pendulum may be possible if simultaneous excitation of both the symmetric and anti-symmetric modes can be achieved.

In this work we neglected (i) the van der Waals force by assuming that it is negligible and (ii) the electrostatic force by assuming that it is distinguishable from the optical force as discussed in [1]. Thermal effects due to material absorption [23] of the fixed laser lines were also not considered. For the optical power used in our model we expect maximum thermal red shifts of around 10 GHz for all cavity modes. This is small compared to the total frequency splitting and is not anticipated to affect the stability of the trap. Finally, we note that trapping limitations may arise from using different sized microspheres in the experimental arrangement, since it would be almost impossible to fabricate identical microspheres. While theoretically [1] is has been shown that a reasonable degree of size mismatch is permissible, it is worth noting that high Q mode splitting using spheres that were dramatically size mismatched (240 $\mu$m and 375 $\mu$m) has been demonstrated experimentally [15]. We, therefore, conclude that size mismatching during the fabrication process should not significantly reduce the trapping effects. If the technical challenges can be overcome, optical corralling of micro-optic components opens up ways to construct optomechanical devices with extremely well-controlled position of the mechanical elements.

This work was supported in part by Science Foundation Ireland under Grant Nos. 06/W.1/I866 and 07/RFP/PHYF518. YW acknowledges support from IRCSET under the Embark Initiative.

FIGURE CAPTIONS

FIG. 1 (color online). Proposed experimental setup. Sphere 1 is stationary and optically coupled to a tapered optical fiber, while Sphere 2 is free to move and optically coupled to Sphere 1. The pumping laser field consists of two monochromatic field components with frequencies $v_1$ and $v_2$. Insert shows the transmission of a weak probe signal as a function of frequency for a given separation, $x$, between the microspheres.

FIG. 2 (color online). (a) Qualitative dependence of the frequencies of the symmetric (lower line) and anti-symmetric (upper line) optical mode in a two-sphere system as a function of the sphere separation; (b) nature of the optical force on Sphere 2 as a function of sphere separation; (c) shape of the optical potential for Sphere 2 as a function of sphere separation.

FIG. 3 (color online). Photonic molecule system with sphere radii $R = 12.5\ \mu\text{m}$, $\gamma/2\pi = 2\ \text{MHz}$, $n_s = 1.46$, $Q = 10^8$, $\lambda = 1568\ \text{nm}$ and $P = 1\ \text{mW}$. (a) Frequency splitting of the anti-symmetric and symmetric optical modes for an unshifted modal frequency, $v_0 = c/\lambda = 191\ \text{THz}$, as a function of sphere separation. The equilibrium position of Sphere 2 is $x_0 = 250\ \text{nm}$ and the applied laser frequencies, $v_1$ and $v_2$, are blue-detuned from the equilibrium position frequencies, $v_{0a}$ and $v_{0s}$ by $5\gamma$. The arrows show the direction of the partial forces, $F_a$ and $F_s$. (b) Position dependence of the optical force on Sphere 2 for three different detunings of $v_1$ and $v_2$ from $v_{0a}$ and $v_{0s}$. (c) Optical potential for Sphere 2 as a function of position for the same detunings as (b).

**Figure 1**

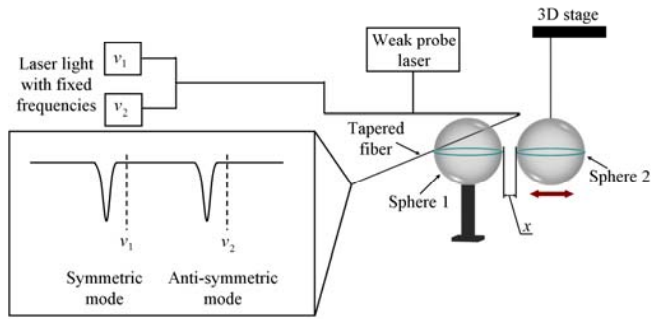

Figure 2

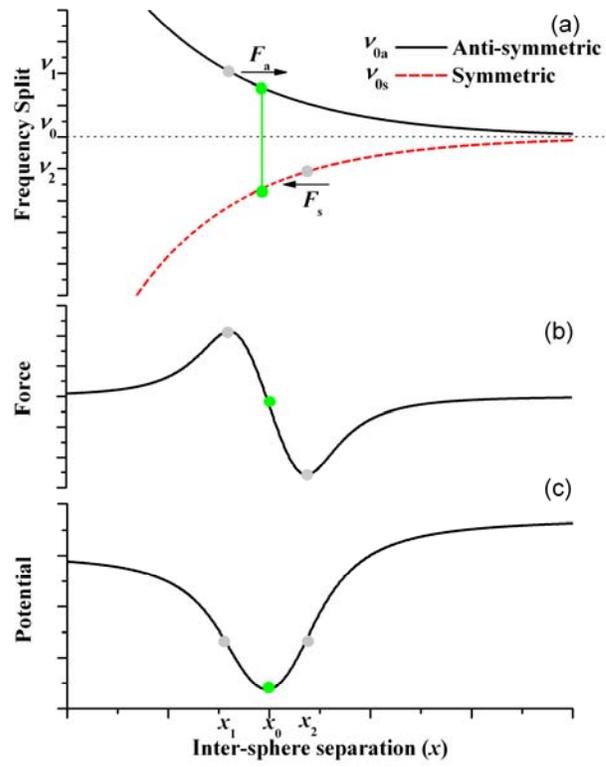

**Figure 3**

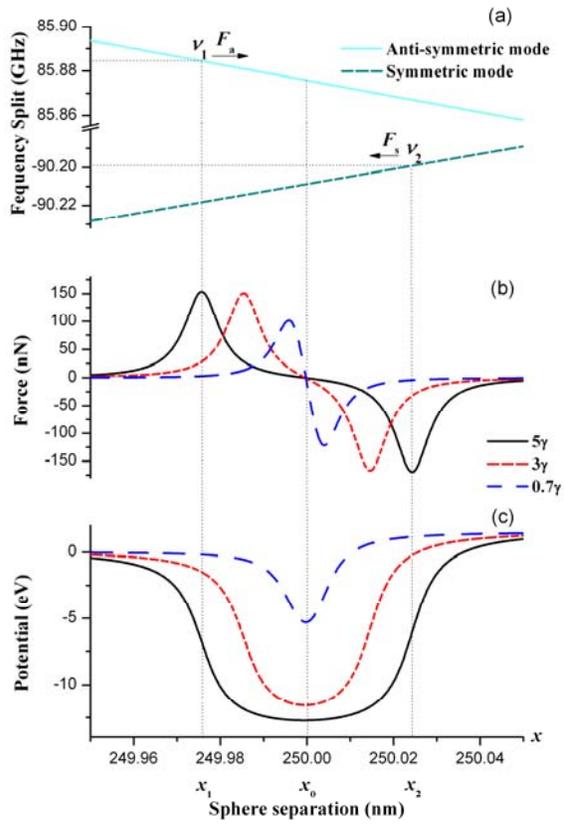